\documentclass[twocolumn,preprintnumbers]{aastex6} 
\usepackage{afterpage}
\usepackage{capt-of} 
\usepackage{diagbox}

\usepackage[figuresright]{rotating}
\usepackage{amsmath}
\usepackage{amssymb,amsmath,latexsym,graphics, graphicx,epsfig,multirow,comment,hyperref,appendix,feyn,slashed,xcolor,afterpage, makecell} 
\usepackage{booktabs}
\usepackage{tabularx}

\newcommand {\be} {\begin {equation}}
\newcommand {\ee} {\end {equation}} 

\newcommand {\bes} {\begin {equation*}}
\newcommand {\ees} {\end {equation*}}

\newcolumntype{L}[1]{>{\raggedright\let\newline\\\arraybackslash\hspace{0pt}}m{#1}}
\newcolumntype{C}[1]{>{\centering\let\newline\\\arraybackslash\hspace{0pt}}m{#1}}
\newcolumntype{R}[1]{>{\raggedleft\let\newline\\\arraybackslash\hspace{0pt}}m{#1}}

\usepackage{csvsimple}

\makeatletter
\newcommand\footnoteref[1]{\protected@xdef\@thefnmark{\ref{#1}}\@footnotemark}
\makeatother

\newcommand{\beq}{\begin{equation}}
\newcommand{\eeq}{\end{equation}}

\begin{document}

\title{Comment on the paper \\ \vspace{0.1in}
``On the correlation between the local dark matter and stellar velocities''}

\author{Mariangela Lisanti}
\affil{Department of Physics, Princeton University, Princeton, NJ 08544, USA\vspace{-0.5cm}}

\author{Lina Necib}
\affil{Walter Burke Institute for Theoretical Physics,
California Institute of Technology, Pasadena, CA 91125, USA}

\maketitle

\vspace{-0.5cm}
We briefly comment on the paper ``On the correlation between the local dark matter and stellar velocities''~\citep{Bozorgnia:2018pfa}, which purports to test our previous claims regarding the correlation of stellar and dark matter kinematics~\citep{eris_paper, Necib:2018igl}.  The conclusions of these papers are not accurately presented in \cite{Bozorgnia:2018pfa}, and we thus clarify them  here.


 In the first study of the~\textsc{Eris} simulation, which is the key paper \cite{Bozorgnia:2018pfa} compares to, \cite{eris_paper} stated in the abstract: ``Using the high resolution
\textsc{Eris} simulation [...], we demonstrate that metal-poor stars are indeed effective tracers
for the local, virialized dark matter velocity distribution."  The choice of qualifier ``virialized'' throughout this early work was deliberate.  While we compared the metal-poor stars to the total dark matter distribution (and not the virialized component specifically), we expected the \textsc{Eris} dark matter halo to be largely virialized.  This is because  \textsc{Eris} had an early accretion history, with no major mergers after  $z\sim3$~\citep{Guedes:2011ux}.  


The work of~\cite{Bozorgnia:2018pfa} did not observe a correlation between metal-poor stars and the total local dark matter distribution in the six \textsc{Auriga} halos, a fact which we do not dispute and largely expect for galaxies with more active merger histories than \textsc{Eris}.  Indeed, we observed precisely the same behavior in the \textsc{Fire} simulations~\citep{Necib:2018igl}.  The discrepancy in~\cite{Bozorgnia:2018pfa} can potentially arise if a relevant fraction of the local dark matter is not virialized, or originated from non-luminous subhalos/smooth accretion.  This would need to be directly tested before drawing conclusions regarding our findings in~\cite{eris_paper, Necib:2018igl}.


Because we observed the same discrepancy as~\cite{Bozorgnia:2018pfa} when we first began studying the \textsc{Fire} halos, we performed a more extensive analysis that divided the local dark matter into separate populations based on origin.  As detailed in~\cite{Necib:2018igl}, this allowed us to conclude the following:\vspace{0.1in}

\noindent 1.  The most metal-poor stars ($\text{[Fe/H]} \lesssim -2$) trace the dark matter accreted from the oldest mergers ($z \gtrsim 3$).  This dark matter population appears to be fully relaxed, or virialized.  \vspace{0.1in}

\noindent 2.  More intermediate metallicity stars ($-2 \lesssim \text{[Fe/H]} \lesssim -1$) trace the dark matter accreted from younger mergers ($z \lesssim 3$).  This correspondence holds for dark matter that is in debris flow, but is less robust for streams. \vspace{0.1in}

\noindent 3.  Dark matter originating from non-luminous satellites or smooth accretion is not necessarily traced by stars.  \vspace{0.1in}

\noindent  Again, neither \cite{eris_paper} nor \cite{Necib:2018igl} claimed that metal-poor stars should trace the \emph{total} local dark matter distribution, as suggested by~\cite{Bozorgnia:2018pfa}.  

Characterizing the dark matter distribution near the Solar position is a challenging open problem.  In tackling this question, \cite{Necib:2018igl} took the approach of finding tracers for separate dark matter populations, as defined by origin.   This approach proved fruitful, and it would be interesting to see 
a similar complete analysis of the \textsc{Auriga} simulations in order to discuss the validity of our conclusions.  

\def\bibsection{} 
\bibliographystyle{aasjournal}
\bibliography{comment_bib}

\end{document}